\documentclass[reprint]{revtex4-2}

\usepackage{graphicx}% Include figure files
\usepackage{dcolumn}% Align table columns on decimal point
\usepackage{bm}% bold math
\usepackage{amsfonts}
\usepackage{amsmath}
\usepackage{physics}
\usepackage{hyperref}% add hypertext capabilities
\hypersetup{
    colorlinks=true,
    linkcolor=blue,
    filecolor=magenta,      
    urlcolor=blue,
    citecolor=red,
    pdftitle={Overleaf Example},
    pdfpagemode=FullScreen,
    }

\begin{document}

\preprint{APS/123-QED}

% \title{Fragile Topological Lattice with Potential Flat Bands and Proposed Realization in Polariton Arrays}
\title{Simple realization of a fragile topological lattice with quasi flat-bands in a microcavity array}
% Force line breaks with \\

\author{Yuhui Wang$^{\dagger}$}
\author{Shupeng Xu$^{\dagger}$}%
\author{Liang Feng}
\author{Ritesh Agarwal}
\email{riteshag@seas.upenn.edu}
\affiliation{
Department of Materials Science and Engineering, University of Pennsylvania,\\Philadelphia, 19104, PA, US}

\date{\today}% It is always \today, today,

% \begin{abstract}
% An article usually includes an abstract, a concise summary of the work
% covered at length in the main body of the article. 
% \begin{description}
% Secondary publications and information retrieval purposes.
% You may use the \texttt{description} environment to structure your abstract;
% use the optional argument of the \verb+\item+ command to give the category of each item. 
% \end{description}
% \end{abstract}

\begin{abstract}

    Topological flat bands (TFBs) are increasingly recognized as an important paradigm to study topological effects in the context of strong correlation physics.
    As a representative example, recently it has been theoretically proposed that the topological non-triviality offers a unique contribution to flat-band superconductivity, which can potentially lead to a higher critical temperature of superconductivity phase transition.
    Nevertheless, the topological effects within flat bands in bosonic systems, specifically in the context of Bose-Einstein condensation (BEC), are less explored.
    It has been shown theoretically that non-trivial topological and geometric properties will also have a significant influence in bosonic condensates as well.
    However, potential experimental realizations have not been extensively studied yet.
    In this work, we introduce a simple photonic lattice from coupled Kagome and triangular lattices designed based on topological quantum chemistry theory, which supports topologically nontrivial quasi-flat bands. 
    Besides band representation analysis, the non-triviality of these quasi-flat bands is also confirmed by Wilson loop spectra which exhibit winding features. 
    We further discuss the corresponding experimental realization in a microcavity array for future study supporting the potential extension to condensed exciton-polaritons. 
    Notably, we showed that the inevitable in-plane longitudinal-transverse polarization splitting in optical microcavities will not hinder the construction of topological quasi-flat bands. 
    This work acts as an initial step to experimentally explore the physical consequence of non-trivial topology and quantum geometry in quasi-flat bands in bosonic systems, offering potential channels for its direct observation.

\end{abstract}
\maketitle

Dispersionless bands, \textit{i.e.}, flat bands, are commonly recognized as one of the key paradigms to study physical phenomena based on strong correlation and interactions due to the dominance of interaction energy $U$ over its quenched kinetic energy.
Recently, it has been theoretically proposed that nontrivial topology can be integrated with flat bands and thus induce novel phenomena, of which one significant example is topological flat-band (TFB) superconductivity \cite{huhtinen2022revisiting, peri2021fragile, peotta2015superfluidity, torma2022superconductivity, cao2018unconventional, tian2023evidence}.
The significance of TFBs in superconductivity is manifested by their connection to quantum metric and contribution to superfluid weight or superfluid stiffness $D_S$. 
Conventionally, $D_S$ is proportional to $n_e/m^{*}$, where $n_e$ is the particle density and $m^{*}$ is the electron effective mass \cite{scalapino1993insulator}.
Therefore, in an isolated trivial flat band, $D_S$ vanishes due to the infinite effective mass. 
Whereas non-trivial topology induces a non-zero lower boundary of $D_S$ from band geometric properties, subsequently defines a non-zero Berezinskii-Kosterlitz-Thouless (BKT) temperature via $T_{\mathrm{BKT}} \propto D_S$, which is the critical temperature in two-dimensional superconductivity \cite{huhtinen2022revisiting, peri2021fragile}.
Though the nontrivial term in $D_S$ was initially discussed in Chern flat bands \cite{peotta2015superfluidity}, it has later been extended to flat bands with fragile topology \cite{peri2021fragile}, which is a topological phase that exhibits obstruction to constructing exponentially-localized, symmetric Wannier functions (in short `Wannier obstruction'), yet can be trivialized by the addition of appropriate trivial bands which is distinct from conventional topological bands \cite{po2018fragile}.

For the characterization of the Wannier obstruction, recently-developed topological quantum chemistry (TQC) theory offers a generalized, efficient approach \cite{bradlyn2017topological, cano2018topology, kruthoff2017topological}.
The general idea of the TQC theory is to examine the topological equivalence to a trivial insulator represented by a set of exponentially localized, symmetric Wannier functions, referred to as the atomic limit.
In the TQC theory, band representations (BRs) are utilized to characterize a band or a set of bands \cite{zak1982band}, which are induced from the Wannier functions in the lattice.
Exponentially localized, symmetric Wannier functions at maximal Wyckoff positions will induce elementary band representations (EBRs), as the minimal elements in the set of BRs.
Therefore, the topological characteristic of a set of bands can be identified based on its equivalence to the EBRs: generally speaking, if a set of bands cannot be represented by a direct sum of EBRs with non-negative integer coefficients, the Wannier obstruction or topological non-triviality of the set of bands can be deduced.

Although in fermionic systems many theoretical and experimental attempts have been made to study TFB superconductivity \cite{huhtinen2022revisiting, peotta2015superfluidity, peri2021fragile, xie2020topology, peri2021fragile, cao2018unconventional, tian2023evidence}, the application of TFBs in bosonic systems is less explored.
Pioneering theoretical works \cite{julku2021excitations, julku2021quantum} have mostly focused on the geometric contribution in the excitation spectra of Bose-Einstein condensation (BEC) in trivial flat bands.
The study of the integration of topological non-triviality with flat band physics is still in a relatively early stage especially in the bosonic context \cite{lukin2023unconventional}.
Considering the profound connection between BEC and superconductivity that both are regarded phenomena related to macroscopic phase coherence, it is intuitively expected that the topological non-triviality may induce novel phenomena as well in bosonic systems.
Unlike electronic materials, for experimental realizations in bosonic systems, synthetic lattices are commonly favored, such as optical lattices of cold atoms and photonic or polaritonic meta-materials.
These systems have been demonstrated to be highly controllable and promising candidates for experimental realizations of condensates in a lattice.

An important question is, how can one design an experimentally-accessible bosonic lattice system with TFBs?
To realize flat bands with non-zero Chern numbers generally entails complex coupling settings \cite{kruchkov2022quantum, sun2011nearly}, meanwhile, since the time-reversal symmetry of bosons is fundamentally distinct from that of fermions, the $\mathbb{Z}_2$ topological insulator is also not accessible \cite{PhysRevLett.131.053802}.
In this case, fragile TFBs designed based on crystalline symmetry naturally emerge as a viable option.
In this work, we propose an experimentally-accessible photonic crystal of microcavity that supports fragile TFBs potentially compatible with BEC physics based on TQC theory, as a first step to extend the study of TFBs in fermionic systems to bosonic systems. 
The main text is organized as follows:
the general strategy of constructing fragile TFBs is reviewed and introduced in the following section based on the disconnected decomposable EBR \cite{cano2018topology, cualuguaru2022general}.
To illustrate the working principle of the construction of TFBs, we introduce the formalism in detail via an explicit example of a Hamiltonian based on s-orbital states.
The topological non-triviality of the constructed TFBs is then confirmed by the winding Wilson loop spectra.
Next, the Hamiltonian is re-constructed with linearly polarized modes at each site, as the electromagnetic analog of the electron s-orbital model, where the physical consequence of in-plane polarization splitting is addressed.
Importantly, the site-symmetry of electromagnetic field analog of s-orbital states is completely different from scalar s-orbital states. 
We nonetheless demonstrate that the change in band representations will not destroy the topological properties of the central flat bands in the Hamiltonian based on in-plane polarization s-orbital states.
Lastly, we discuss the potential experimental realization of fragile TFBs in a microcavity array with detailed numerical simulations, including eigen-spectra and symmetry indicators which are in a nice agreement with our Hamiltonian of linearly polarized modes.

\begin{figure}
    \centering
    \includegraphics[width = 0.45 \textwidth]{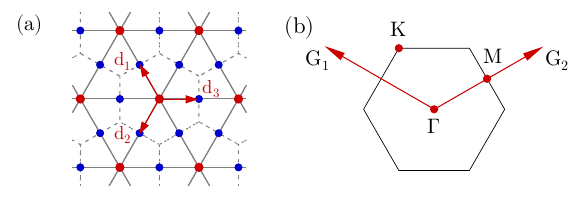}
    \caption{(a) Wyckoff positions 1a (red dots at unit-cell centers) and 3c (blue dots at edge centers) in the space group $p6mm.1’$, with site-symmetry of point group (PG) $6mm$ and $2mm$, forming a triangular and Kagome lattice, respectively.
    (b) High symmetry points in the first Brillouin zone $\Gamma$, $K$ and $M$. Two reciprocal vectors are noted as $\mathrm{G}_1$ and $\mathrm{G}_2$.}
    \label{fig:fig1}
\end{figure}

\begin{figure}
    \centering
    \includegraphics[width = 0.5 \textwidth]{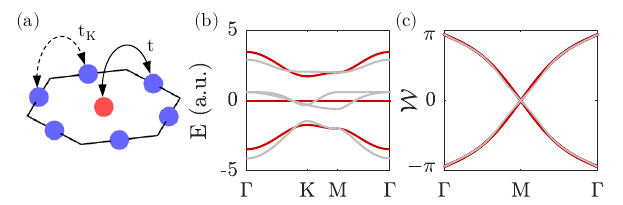}
    \caption{Tight binding model of s-orbital states in $A_1$ representation. (a) Schematic of the lattice, where $t$ measures the coupling between 1a and 3c site (solid arrows), and $t_K$ for the Kagome lattice couplings (dashed arrows). 
    (b) Calculated band diagram with $t = -1.0$, $t_K = 0$ (red) and $t_K = -0.3$ (gray). 
    Note that Kagome coupling renders the central two bands dispersive and non-degenerate. 
    (c) Calculated Wilson loop spectra correspond to $t_K = 0$ (red) and $t_K = -0.3$ (gray), both showing a winding feature.}
    \label{fig:fig2}
\end{figure}

\textit{Principle of construction of fragile TFBs.---}
It has been realized from the previous work \cite{PhysRevLett.131.053802, de2019engineering} that photonic crystals of the space group $p6mm.1'$ potentially support the fragile topological phase if appropriately disconnected.
This intuition reflects the general statement that fragile topological phases can be derived from disconnecting a decomposable EBR \cite{cano2018topology}, as a specific case.
A decomposable EBR, $\mathcal{B}_0$, can be split into two quasiband representations, obeying compatibility relations: $\mathcal{B}_0 = \mathcal{B}_1 \oplus \mathcal{B}_2$.
By definition, $\mathcal{B}_1$ and $\mathcal{B}_2$ cannot be EBRs simultaneously; nevertheless, they can represent two sets of stable or fragile topological bands, or represent one set of fragile topological bands and one EBR if disconnected \cite{po2018fragile}.
For the latter case, without loss of generality, we let $\mathcal{B}_1$ be an EBR, therefore $\mathcal{B}_2$ necessarily represents a set of bands with non-trivial fragile topology.
In the current $p6mm.1'$ scenario, since s-orbital states at 3c (point group $mm2$) Wyckoff positions can be represented by the irreducible representation (irrep) $A_1$ (Table. \ref{tab:mm2reps}), the corresponding BR can be written as:
\begin{equation}
\begin{aligned}
    (A_1 \uparrow G)_{3c} &= \{\Gamma_1 \oplus \Gamma_5, K_1 \oplus K_3, M_1 \oplus M_3 \oplus M_4\}\\
    &= \{\Gamma_1, K_1, M_1 \} \oplus \{ \Gamma_5, K_3, M_3 \oplus M_4 \},
    \end{aligned}
\end{equation}
where $\Gamma$, $K$ and $M$ are high symmetry momenta marked in Fig. \ref{fig:fig1}b.
Notice that $\{\Gamma_1, K_1, M_1\}$ is an EBR and can be induced from s-orbital states at 1a site (point group $mm6$), represented by $A_1$ (Table. \ref{tab:6mmreps}).
Hence according to the previous deduction the remaining two bands $\{ \Gamma_5, K_3, M_3 \oplus M_4 \}$ are a set of fragile topological bands.
In the Kagome lattice, $(A_1 \uparrow G)_{1a}$ can be constructed as the quasi-s-orbital state formed by the equal-phase superposition of s-orbital states at 3c sites.
Naturally, we can introduce another s-orbital state at 1a site, corresponding to a triangular lattice (red sites in Fig. \ref{fig:fig1}a), to only couple with the quasi-s-orbital state.
This coupling setting can be realized by introducing identical real couplings only between 1a and nearest neighboring 3c sites (solid red arrow in Fig. \ref{fig:fig2}a) preserving $C_6$ symmetry.
If all couplings are set to zero, the lattice is at its atomic limit, naturally forming four (1+3) degenerate flat bands. 
After introducing these couplings between 1a and 3c sites, two complete band gaps will be formed across the entire BZ, and the residual two bands will stay degenerate and flat with the representation:
\begin{equation}
    \{A_1 \uparrow G \}_{3c} \ominus \{A_1 \uparrow G \}_{1a} =\{\Gamma_5,K_3,M_3 \oplus M_4 \}.
    \label{eqn:br}
\end{equation}
Consequently, we construct two fragile topological flat bands based on the simple couplings between two lattices.

\begin{table}
\setlength{\tabcolsep}{15pt}
\renewcommand{\arraystretch}{0.9}
\centering
\caption{\label{tab:mm2reps} \textbf{Irreducible representations $A_1$, $B_1$, $B_2$ of the point group $mm2$.} Detailed description of symmetry operations can be found in Bilbao Crystallographic Server \cite{aroyo2006bilbao1, aroyo2006bilbao2, aroyo2011crystallography}.}
% \begin{ruledtabular}
\begin{tabular}{ c c c c c }
\hline
\textrm{Reps.}&
\textrm{$m_{100}$}&
\multicolumn{1}{c}{$m_{010}$}&
\textrm{$2_{001}$}\\

\hline
$A_1$ & 1 & 1 & 1\\
$B_1$ & -1 & 1 & -1\\
$B_2$ & 1  & -1 & -1\\
\hline
\end{tabular}
% \end{ruledtabular}
\end{table}

\begin{table}
\setlength{\tabcolsep}{15pt}
\renewcommand{\arraystretch}{0.9}
\centering
\caption{\label{tab:6mmreps} \textbf{Irreducible representations $A_1$, $E_1$ of the point group $6mm$.} $\mathrm{M_1}$ stands for mirror symmetry operations $m_{210}$, $m_{1\Bar{1}0}$ and $m_{120}$. $\mathrm{M_2}$ for $m_{100}$, $m_{110}$ and $m_{010}$.
Detailed description of symmetry operations can be found in Bilbao Crystallographic Server \cite{aroyo2006bilbao1, aroyo2006bilbao2, aroyo2011crystallography}.}
% \begin{ruledtabular}
\begin{tabular}{ c c c c c }
\hline
\textrm{Reps.}&
\textrm{$M_1$}&
\multicolumn{1}{c}{$M_2$}&
\textrm{$2_{001}$}\\

\hline
$A_1$ & 1  & 1  & 1\\
% $B_1$ & -1 & 1 & -1\\
% $B_2$ & 1  & -1 & -1\\
$E_1$ & 0  & 0 & -2\\
\hline
\end{tabular}
% \end{ruledtabular}
\end{table}

Subsequently, we explicitly write down the Hamiltonian and take other possible degrees of freedom into consideration.
With two distinct on-site energies at 1a and 3c sites and the couplings among 3c (Kagome) sites included, the total tight-binding Hamiltonian as the sum of three parts is:
\begin{equation}
\begin{aligned}
    \mathcal{H} &=  2 t\sum_{i = 1}^3 [a_{1a}^{\dagger}a_{3c,i}\cos(k_i) + h.c.],\\
    \mathcal{H}_0 &= \frac{U}{2} (a_{1a}^{\dagger}a_{1a} - \sum_{i=1}^{3}a_{3c, i}^{\dagger}a_{3c,i}),\\
    \mathcal{H}_{\mathrm{K}} &= 2t_K \sum_{i,j} [a_{3c,i}^{\dagger} a_{3c,j}\cos(k_i - k_j) + h.c.].
\end{aligned}
\end{equation}
$a_{1a,3c}$ is the annihilation operator of orbital states of corresponding Wyckoff positions.
$\mathcal{H}$ is the Hamiltonian that captures the core idea of the construction of fragile TFBs that orbitals at 1a site couple with orbitals at 3c sites with a real and constant coupling strength $t$.
$\mathcal{H}_0$ and $\mathcal{H}_{\mathrm{K}}$ address the effect of distinct onsite energy $\pm U/2$ and coupling between 3c and 3c sites, \textit{i.e.}, Kagome coupling, given by $t_K$.
In $\mathcal{H}$ and $\mathcal{H}_K$, $k_i  = k \cdot d_i$, where $d_1  = a [ -1/2,\sqrt{3}/2 ] $, $d_2=a[-1/2,-\sqrt{3}/2]$ and $d_3=a[1,0]$, and $a$ is the distance between nearest neighboring sites (Fig. \ref{fig:fig1}a).
$h.c.$ corresponds to Hermitian conjugate. 

The most simplified, ideal case is with $U=0$ and $t_K=0$, of which we only need to discuss $\mathcal{H}$.
The corresponding band diagram is shown in red in Fig. \ref{fig:fig2}b, where two degenerate perfectly flat bands at zero energy can be observed. 
The physical picture is also nicely presented in the molecular-orbital (MO) representations: $\mathcal{H} = 2 t( M_1^{\dagger}M_2 + h.c.)$, where the two molecular orbital operators are defined as $M_1 = a_{1a}$ and $M_2 = \sum_{i=1}^{3}a_{3c,i}\cos(k_i)$ \cite{mizoguchi2019molecular}, identical to our previous deduction.
Writing $\mathcal{H}$ in the MO representation has two important implications:
first, the orbital state at 1a is introduced to couple with one molecular orbital formed by 3c orbital states, which is the quasi-s-orbital states we just discussed, written as $(A_1 \uparrow G)_{1a}$.
Second, from the MO representation, it is elucidated that the rank of the Hamiltonian is 2 though four orbital states are originally involved to define the Hamiltonian, which explains the origin of the perfect band flatness.
In fact, it also suggests that the difference of on-site energy at 1a and 3c sites will not affect the topological non-triviality and band flatness, since one can always rewrite $\mathcal{H}_0 = \frac{U}{2} (a_{1a}^{\dagger}a_{1a} - \sum_{i=1}^{3}a_{3c, i}^{\dagger}a_{3c,i})$ to $\mathcal{H}_0 = U a_{1a}^{\dagger}a_{1a} = U M_1^{\dagger}M_1$.
This MO representation will not be valid if non-zero Kagome couplings $t_K$ are included, hence the central two bands are not strictly flat and degenerate with a non-zero $t_K$ (see gray bands in Fig. \ref{fig:fig2}b with $t_K = -0.3$). 
However, the non-trivial fragile topology is preserved as long as the band representation of the central two bands is not altered by $t_K$.
The bands will be connected at $t = t_K$ thus as long as the magnitude of $t_K$ is smaller than $t$ the current band representations (Eqn. \ref{eqn:br}) are valid.
Another indicator of the topological non-triviality is the winding Wilson loop spectrum. 
Wilson loop is calculated as \cite{alexandradinata2014wilson},
 \begin{equation}
     \mathcal{W}_C=\mathcal{P}\exp\left[ i\oint_C \mathbf{A}(\mathbf{k})\cdot d\mathbf{k} \right]
     \label{eqn:wilson1}
 \end{equation}
where 
\begin{equation}
[\mathbf{A} (\mathbf{k})]_{mn}=i\bra{u_m(\mathbf{k})}\nabla_\mathbf{k}\ket{u_n(\mathbf{k})}
\label{eqn:wilson2}
\end{equation}
is the non-Abelian Berry connection for the composite bands indexed by $m$ and $n$. 
The geometry of the Wilson loop calculation is denoted in Fig. \ref{fig:fig1}b, where the closed loop $C$ is along the reciprocal lattice vector $G_1$ and the spectra are plotted as the loop moves along $G_2$ following the $\Gamma$ – M – $\Gamma$ path. 
The spectrum of the calculated Wilson loop indicates a winding feature providing direct evidence that the central two bands of the proposed system are Wannier obstructed and hence topologically nontrivial (red in Fig. \ref{fig:fig2}c).
Another calculated Wilson loop spectrum corresponds to the $t_K = -0.3$ case shown in Fig. \ref{fig:fig2}b, confirming that weak Kagome coupling preserving the band representations will not trivialize the central two bands (gray in Fig. \ref{fig:fig2}c).

\textit{Tight-binding model of electromagnetic modes.---}
In this section, we discuss the electromagnetic (EM) field realization of the TFBs, extended from our previous construction.
In order to explore the physical consequence of the topological non-triviality in a bosonic system beyond the single particle picture, we specifically consider the EM realization in a microcavity array due to its compability to exciton-polariton condensates which enable direct observation across both real and momentum spaces, including dispersion, excitation spectra, and phase coherence \cite{byrnes2014exciton, baboux2016bosonic, amo2009superfluidity, pieczarka2020observation}.

\begin{figure}
    \centering
    \includegraphics[width = 0.5 \textwidth]{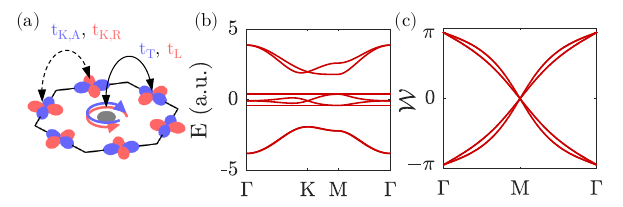}
    \caption{Tight binding model of the EM-field realization of TFBs in $E_1$, $B_1$ and $B_2$ irreps. 
    (a) Schematic of the lattice, $t_T$ stands for the transverse coupling to the azimuthal modes (blue) and $t_L$ stands for the longitudinal coupling to the radial modes (red). 
$t_{K,A}$ stands for the couplings between the azimuthal modes at 3c Kagome sites, and $t_{K,R}$  stands for the Kagome couplings as well but between radial modes.    
    (b) Calculated band diagram with $t_L = -1.2$, $t_T = -1$, $U_A = 0.4$, $U_R = -0.4$ and $U_{1a} = 0$ with vanished Kagome couplings. 
    Two flat bands can be noticed near zero at $U_A$ and $U_R$. 
    (c) Corresponding Wilson loop of the central four bands, where four branches can be observed, and both sets exhibit a winding feature.}
    \label{fig:fig3}
\end{figure}

Due to the finite longitudinal-transverse (LT) polarization splitting \cite{nalitov2015spin}, which is an in-plane version of transverse-electric–transverse-magnetic (TE-TM) splitting, originating from distinct boundary conditions for TE and TM modes in Maxwell equations, two orthogonal polarized modes need to be taken into consideration independently at each site.
For these in-plane polarization modes in the microcavity system, the $C_2$ eigenvalue of the analog s-orbital states is $-1$, distinct from electronic s-orbital states, and site-symmetry representations therefore change accordingly in the EM realization.
The site-symmetry at 1a Wyckoff positions remains PG $6mm1’$ and we choose the irrep to be $E_1$ since there is no symmetry breaking that breaks the degeneracy of two circularly polarized modes.
The site-symmetry PG $2mm.1’$ at 3c Wyckoff positions supports two independent and orthogonal linearly polarized states: one is parallel to edges of the unit cell and the other one is normal to edges, noted as azimuthal (A) and radial (R) modes and corresponding to $B_1$ and $B_2$ representations, respectively. 
Azimuthal and radial modes are represented in p-orbital-like features in Fig. \ref{fig:fig3}a, color-coded in blue and red, respectively, to illustrate how they are aligned.
We emphasize that no modes with higher orbital angular momentum numbers are involved in the current model.

The new induced band representations accounting for eigenmodes corresponding to different polarizations can be written as:
\begin{equation}
\begin{aligned}
    (E_1 \uparrow G)_{1a} &= \{\Gamma_6, K_3, M_3 \oplus M_4\},\\
    (B_1 \uparrow G)_{3c} &= \{\Gamma_3 \oplus \Gamma_6, K_1 \oplus K_3, M_1 \oplus M_2 \oplus M_3\},\\
    (B_2 \uparrow G)_{3c} &= \{\Gamma_4 \oplus \Gamma_6, K_2 \oplus K_3, M_1 \oplus M_2 \oplus M_4\}.
\end{aligned}
\end{equation}
Similarly, we obtain the expected fragile topological band representation:
\begin{equation}
    \begin{aligned}
        (&B_1 \uparrow G)_{3c} \oplus (B_2 \uparrow G)_{3c} \ominus (E_1 \uparrow G)_{1a} =\\
    \{&\Gamma_3 \oplus \Gamma_4 \oplus \Gamma_6, K_1 \oplus K_2 \oplus K_3, 2(M_1 \oplus M_2)\},
    \end{aligned}
    \label{eqn:rep}
\end{equation}
which is not an EBR.
Note that in Eqn. \ref{eqn:rep}, the $C_2$ symmetry eigenvalues of co-representations at $\Gamma$, namely $\Gamma_{3,4,6}$, are $-1$ and those of co-representations $M_{1,2}$ are $+1$, respectively.
This specific combination of symmetry eigenvalues cannot be found in any EBR in $p6mm.1'$ group, hence can be used as a quick symmetry indicator of topological bands.
The main parameters in the corresponding Hamiltonian include on-site energy of two degenerate modes at 1a site $U_{1a}$ and on-site energy of azimuthal and radial modes at 3c sites $U_A$ and $U_R$, respectively;
The coupling between azimuthal and radial modes to 1a modes are noted as $t_T$ and $t_L$, respectively. 
To account for the effect of Kagome couplings between 3c sites, we also introduce the Kagome couplings between azimuthal modes and radial modes to be $t_{K,A}$ and $t_{K,R}$, respectively.
Detailed Hamiltonian representing the EM field case is discussed in supplementary materials.

The band diagram calculated from this EM tight-binding model is shown in Fig. \ref{fig:fig3}b, with $t_L = -1.2$, $t_T = -1$, $U_A = 0.4$, $U_R = -0.4$. 
We intentionally defined $t_T \neq t_L$ and $U_A \neq U_R$ so that the central four bands are not completely degenerate.
With this setup, two upper bands, four central bands and two lower bands, correspond to band representations $(E_1 \uparrow G)_{1a}$, 
$(B_1 \uparrow G)_{3c} \oplus (B_2 \uparrow G)_{3c} \ominus (E_1 \uparrow G)_{1a}$, 
and $(E_1 \uparrow G)_{1a}$, respectively, as designed (Eqn. \ref{eqn:rep}). 
With zero Kagome coupling in the model, two perfectly flat bands can be observed in Fig. \ref{fig:fig3}b.
Specifically, these two bands contain only azimuthal or radial modes at 3c sites, which is due to the symmetry constraint of the higher synthetic orbital states in 3c site coupling to 1a site, similar to the s-orbital scenario.
Therefore, azimuthal and radial modes will not mix and necessarily remain at their original polarization. 
The corresponding Wilson loop of the central four bands is demonstrated in Fig. \ref{fig:fig3}c, indicating two sets of winding branches in the Wilson spectra, degenerate at 0 and $\pm \pi$ points.

\textit{Electromagnetic numerical simulations.---}
To further examine the experimental accessibility and evaluate the performance of our theoretical model, we carried out 3D EM numerical simulations of the eigenfrequency spectrum of a planar microcavity system consisting of etched micro-resonators, functioning as synthetic atoms in the lattice, similar to Refs. \cite{jacqmin2014direct, nalitov2015polariton, nalitov2015spin}.
The detailed parameters can be found in the Supplementary Materials.

\begin{figure}[t!]
    \centering
    \includegraphics[width = 0.5 \textwidth]{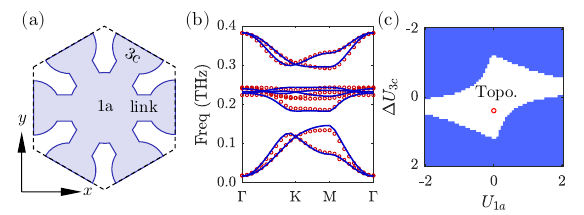}
    \caption{(a) Schematic of the planar cavity lattice structure in numerical simulations.
    The cavity region corresponding to high refractive index are shaded.
    Cylindrical resonators at 1a and 3c sites are connected by links, while 
    resonators at 3c sites are mutually separated.
    (b) Numerically simulated band diagram (marked in red circles) and tight binding fitting (plotted in blue lines). 
    A complete gap can be observed between upper two and central four bands.
    A base frequency of 303.95 THz has been subtracted from the frequency spectrum.
    (c) Parameter phase diagram predict by the tight binding model, where we define the  $t_T = -0.4$, $t_L = -0.7$, $t_{K,A} = -0.02$, $t_{K,R} = 0.1$ and scan $\Delta U_{3c}$ and $U_{1a}$.
    The central white regime corresponds to the fragile topological phase of the lattice. 
    The parameter set in (a) falls into the topological phase after scaling, marked as a red circle.}
    \label{fig:fig4}
\end{figure}

To map the proposed electromagnetic tight-binding model to this structure, we intentionally introduce an extended spacing to reduce (ideally to vanish) the overlapping of EM fields between each resonator, and subsequently only bridge resonators at 1a and 3c sites to locally enhance couplings between them via links to enhance the coupling contrast (Fig. \ref{fig:fig4}a).
These links will consequently change the shape of resonators at 3c Wyckoff positions, hence inevitably affecting the on-site energy of azimuthal and radial modes.
The elliptical aspect ratio of the resonators at 3c sites is tuned accordingly whereas since links connected to 1a site will not break the degeneracy of two circularly polarized modes, it is not necessary to change the shape of resonators at 1a Wyckoff positions. 
The elliptical design preserves the space group without breaking any crystalline symmetry of the structure and hence corresponding induced band representations are still valid. 
The detailed geometric parameters are listed in supplementary materials.

The side boundaries of the unit cell are set to be Bloch boundary conditions, whereas perfect electric conductor (PEC) boundaries are assigned at the top and bottom of the cavity to set up ideal micro-cavities instead of computationally costly distributed Bragg reflector (DBR) structures.
The simplified PEC boundary condition is first validated as an effective approximation to capture the LT splitting phenomenon:
we numerically simulated a typical photonic microcavity graphene lattice with the PEC approximation, and the results of band dispersion of the photonic graphene are compared to Ref. \cite{nalitov2015spin}.
(Detailed comparison can be found in the supplementary materials.)
The central wavelength of the cavity structure is set to be 1000 nm and the refractive index of the cavity is set to be 4 in order to further suppress the Kagome couplings.
The high refractive index is an assumption that exciton-polaritons will have a higher effective mass compared to its photonic counterparts due to the large mass of excitons.

We numerically simulated the eigenfrequency spectrum of the structure and it was then fitted based on the tight binding model discussed in the previous section (red circles and blue lines in Fig. \ref{fig:fig4}b, respectively).
The fitting parameters for the numerical simulation results are
$t_{L} = -64.6$ GHz, $t_{T} = -36.1 $ GHz, 
$t_{K,A} = -1.9$ THz, $t_{K,R} = 9.5$ GHz,
$U_A = 38$ GHz, $U_R = 0.0$ GHz and $U_{1a} = 0.0$ GHz.
Note that fitting parameters are now assigned with units to capture the actual energy scale in practice.
In the current results, band representations and degenracies are correctly captured.
Regardless of minor disagreements in Fig. \ref{fig:fig4}b, the numerical results indeed agree with the electromagnetic tight-binding model.
Four eigenmodes at $M$ point exhibiting $C_2$ eigenvalues $+1$, which captures the main feature of topological non-triviality.
The upper bands of the central four bands show the flatness of $\sim 1/10$ of gap width.

With the tight-binding model demonstrated to be capable of capturing the correct band representations in microcavity arrays, we then illustrate the calculated fragile topological phase diagram based on the tight-binding model in the parameter space (Fig. \ref{fig:fig4}c), which can be helpful for further optimization of the structural design of a TFB microcavity array.
Based on the fitting parameters of the simulation results in Fig. \ref{fig:fig4}a, we set $t_T = -0.38$, $t_L = -0.68$, with finite Kagome coupling $t_{K,A} = -0.02$, $t_{K,R} = 0.1$, $U_R = 0$, and scanning $\Delta U_{3c} =U_A - U_R$ and $U_{1a}$ from $-2$ to $2$.
The topological phase is numerically characterized by two features: (1) the presence of two complete gaps between the central four bands with the upper and lower two bands; and (2) $C_2$ eigenvalues at the $M$ point of the central four bands are $+1$.
The central region in Fig. \ref{fig:fig4}c corresponds to the fragile topological phase, confirming our intuitive conjecture that a suppressed $\Delta U_{3c}$ and $U_{1a}$ will be beneficial for realizing fragile topological phases.
Additionally, we deduce from our previous calculation (Fig. \ref{fig:fig3}b) that Kagome coupling is the most significant ingredient for enhanced band flatness.
However, due to the non-negligible overlapping of electromagnetic modes in the microcavity array, it is challenging to selectively tune off the Kagome coupling.
We expect that computer-assisted inverse design may be applied to further optimize the structure, and our work has hence offered a significant starting point.

In summary, we exploited the TQC theory to design a simple photonic lattice with fragile topological quasi-flat bands, completely isolated from other trivial bands, manifesting its Wannier obstructed properties.
We further proposed a microcavity array realization, where LT polarization splitting is necessarily taken into consideration, that supports these fragile quasi-topological topological bands in the context as well though band representations are necessarily updated. 
A tight binding description and numerical simulations of the proposed microcavity system together demonstrate the validity of this topological lattice and can estimate the parametric requirement for future optimization and applications.
Our designs are compatible with coupling the photonic cavity to excitonic materials to enable new experimental observations of the
effect of quantum geometry and topology on polariton condensates \cite{liu2020generation, huhtinen2024quantum}.

$^{\dagger}$: These authors contributed equally to this work.
\begin{acknowledgments}
This work was supported by the Office of Naval Research via grant No.N00014-22-1-2378.
\end{acknowledgments}

\bibliography{apssamp}% Produces the bibliography via BibTeX.

\end{document}